\documentclass[journal=nalefd,manuscript=letter,layout=twocolumn,]{achemso}
\usepackage[sc]{mathpazo}
\usepackage[scaled=0.9]{helvet}
\usepackage[utf8]{inputenc}
\usepackage{graphicx}
\usepackage{amsmath,amssymb,amsfonts}
\usepackage{upgreek} 
\usepackage{textcomp}
\usepackage[pdftex,dvipsnames]{xcolor}
\usepackage[pdftex]{hyperref}
\hypersetup{
	pdftitle={Laehnemann_Radial-Stark-effect-in-InGaN-nanowires},
	colorlinks,
	citecolor=blue
	}
\usepackage{soul}
\usepackage{xspace}

\newcommand{\celsius}{$^{\circ}$C\xspace}

\title{Radial Stark effect in (In,Ga)N nanowires}


\author{Jonas Lähnemann}
\altaffiliation{Present address: Equipe mixte CEA-CNRS-UJF Nanophysique et Semiconducteurs, INAC/SP2M, CEA-Grenoble, 17 rue des Martyrs, 38054 Grenoble, France}
\author{Pierre Corfdir}
\email{corfdir@pdi-berlin.de}
\altaffiliation{J. Lähnemann and P. Corfdir have contributed equally to this work.}
\author{Felix Feix}
\author{Jumpei Kamimura}
\author{Timur Flissikowski}
\author{Holger T. Grahn}
\author{Lutz Geelhaar}
\author{Oliver Brandt}
\affiliation{Paul-Drude-Institut für Festkörperelektronik, Hausvogteiplatz 5--7, 10117 Berlin, Germany}

\begin{document}


\begin{abstract}

We study the luminescence of unintentionally doped and Si-doped In$_x$Ga$_{1-x}$N nanowires with a low In content ($x < 0.2$) grown by molecular beam epitaxy on Si substrates. The emission band observed at 300~K from the unintentionally doped samples is centered at much lower energies (800~meV) than expected from the In content measured by x-ray diffractometry and energy dispersive x-ray spectroscopy. This discrepancy arises from the pinning of the Fermi level at the sidewalls of the nanowires, which gives rise to strong radial built-in electric fields. The combination of the built-in electric fields with the compositional fluctuations inherent to (In,Ga)N alloys induces a competition between spatially direct and indirect recombination channels. At elevated temperatures, electrons at the core of the nanowire recombine with holes close to the surface, and the emission from unintentionally doped nanowires exhibits a Stark shift of several hundreds of meV. The competition between spatially direct and indirect transitions is analyzed as a function of temperature for samples with various Si concentrations. We propose that the radial Stark effect is responsible for the broadband absorption of (In,Ga)N nanowires across the entire visible range, which makes these nanostructures a promising platform for solar energy applications.

\end{abstract}

\begin{table*}
\caption{Summary of the most important properties of the samples under investigation. Shown are the substrate temperature $T_\text{sub}$, the temperature of the Si effusion cell $T_\text{Si}$, the mean equivalent disk diameter $\langle d_\text{disk}\rangle$ of the nanowires and its standard deviation, the average length $L$ of the nanowires, their coalescence degree $\sigma_C$, and their average In content $x$. For all samples, $x$ has been deduced from both x-ray diffractometry (XRD) and energy-dispersive x-ray spectroscopy (EDX).}
\begin{tabular}{llccccc}
\hline\hline
$T_\text{sub}$ (\celsius)& $T_\text{Si}$ (\celsius)& $\langle d_\text{disk}\rangle$ (nm) & $L$ (nm) & $\sigma_C$ & $x_\text{XRD}$ & $x_\text{EDX}$ \\
\hline
590 & stand-by & $115\pm93$ & 270 & 0.97 & 0.16 & 0.20\\
640 & stand-by & $91\pm68$ & 270 & 0.92 & 0.06 & 0.10\\
640 & 1200 & $70\pm40$ & 270 & 0.82 & 0.04 & 0.07\\
640 & 1250 & $58\pm28$ & 270 & 0.75 & 0.02 & 0.04\\
640 & 1300 & $42\pm13$ & 490 & 0.66 & 0.02 & 0.02\\
\hline\hline
\end{tabular}
\label{Table1}
\end{table*}

\section{}

The possibility to tune the bandgap of (In,Ga)N across the whole visible spectral range makes this ternary alloy highly attractive for solar harvesting applications such as solar cells or photoelectrochemical water splitting.\cite{Jani2007,Hwang2012,Kamimura2013,Alotaibi_nl_2013,Kibria_acsnano_2013,Caccamo_aami_2014,Ebaid2015}. However, strain relaxation in planar (In,Ga)N/GaN heterostructures occurs through the generation of extended nonradiative defects, which are detrimental for optoelectronic devices. The growth of (In,Ga)N in the form of nanowires lifts this constraint, since strain can relax at the nanowire sidewalls.\cite{Gudiksen2002} In addition, the nanowire geometry is beneficial regarding an efficient coupling between light and matter, and enhanced conversion efficiencies have for instance been reported for III-V semiconductor nanowire solar cells.\cite{Wallentin2013,Krogstrup2013}

Another peculiarity of wurtzite group-III-nitride heterostructures grown along the polar \textit{c} direction arises from the large differences in the spontaneous and  piezoelectric polarizations at the heterostructure interfaces. These discontinuities in the polarization field give rise to built-in electric fields, whose magnitude is on the order of MV/cm in a quantum well. As a result, electrons and holes are localized at opposite interfaces of the quantum well, and the associated redshift of the transition energy and the reduction in the overlap of electron and hole wavefunctions are commonly referred to as quantum-confined Stark effect.\cite{Miller_prl_1984,Takeuchi_jjap_1997} In group-III nitride nanowires, a similar spatial separation of electrons and holes can result from the radial electric fields that accompany the surface band bending as a consequence of the Fermi level pinning at the lateral surface of the nanowires.\cite{Calarco2005,vanWeert2006,Thunich2009,Reshchikov2009,Calarco2011,Corfdir2014} In particular, in GaN nanowires, this separation can be large enough to significantly reduce the wavefunction overlap and lead to a quenching of the nanowire photoluminescence intensity.\cite{Pfuller2010,Lefebvre2012} The situation becomes even more complex when considering the case of (In,Ga)N nanowires. The experimentally observed presence of radial electric fields\cite{Lefebvre2012,Wallys2012} is accompanied by strong carrier localization.\cite{Chichibu1997,Martin1999} This localization may occur at compositional fluctuations in the ternary alloy\cite{Chichibu1997,Martin1999,Goodman2011,Segura-Ruiz2011,Murotani2013,You2013,Segura-Ruiz2014,Lahnemann_jpd_2014} or at random dopant fluctuations in the nanowire\cite{Marquardt2015} and may result in individual electron and hole localization at random spatial positions.

In this letter, we study experimentally the combined role of radial surface electric fields and carrier localization  in the ternary alloy on the emission properties of spontaneously formed (In,Ga)N nanowires. We show that the interplay of these two effects can give rise to a radial equivalent to the quantum confined Stark effect. Our study is motivated by the discrepancy between the alloy composition measured by x-ray diffractometry and energy-dispersive x-ray spectroscopy on the one hand, and the peak energy of the dominant emission band observed by cathodoluminescence spectroscopy on the other hand. To investigate the origin of this discrepancy, we perform temperature-dependent and time-resolved photoluminescence spectroscopy on both unintentionally doped and Si-doped nanowire ensembles. These investigations reveal that the main emission band in the undoped (In,Ga)N nanowires stems from the recombination of radially separated electrons and holes, which thus experience a radial Stark effect causing a strong red-shift of the corresponding radiative transitions. Furthermore, we highlight the important role of surface states for the magnitude of the radial electric fields by photoluminescence measurements following a dilute hydrochloric acid etch as well as by cathodoluminescence experiments following prolonged electron irradiation.

\begin{figure*}[t]
\centering
\includegraphics*[width=14cm]{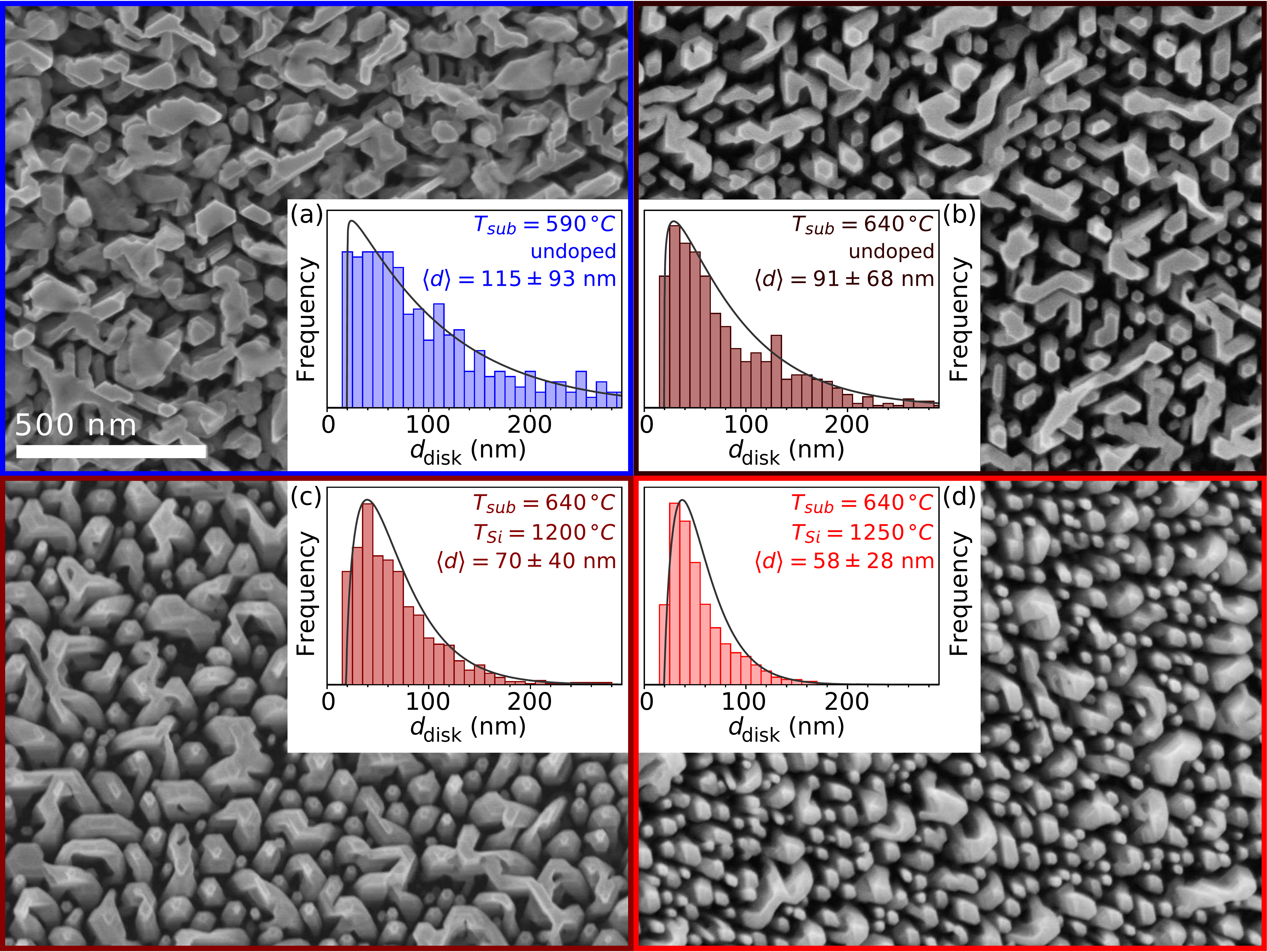}
\caption{\label{fig:sem}Top-view SEM images of the undoped nanowire ensembles for (a) $T_\text{sub} = 590$\,\celsius and (b) $T_\text{sub} = 640$\,\celsius as well as of Si-doped samples grown at $T_\text{sub} = 640$\,\celsius with Si cell temperatures of (c) 1200\,\celsius and (d) 1250\,\celsius. The insets show histograms representing the distribution of equivalent disk diameters ($d_\text{disk}$). The mean values $\langle d_\text{disk} \rangle$ and variances are determined from the fits with shifted Gamma distributions as shown in the figure.}
\end{figure*}

\begin{figure*}[t!]
\centering
\includegraphics*[width=14cm]{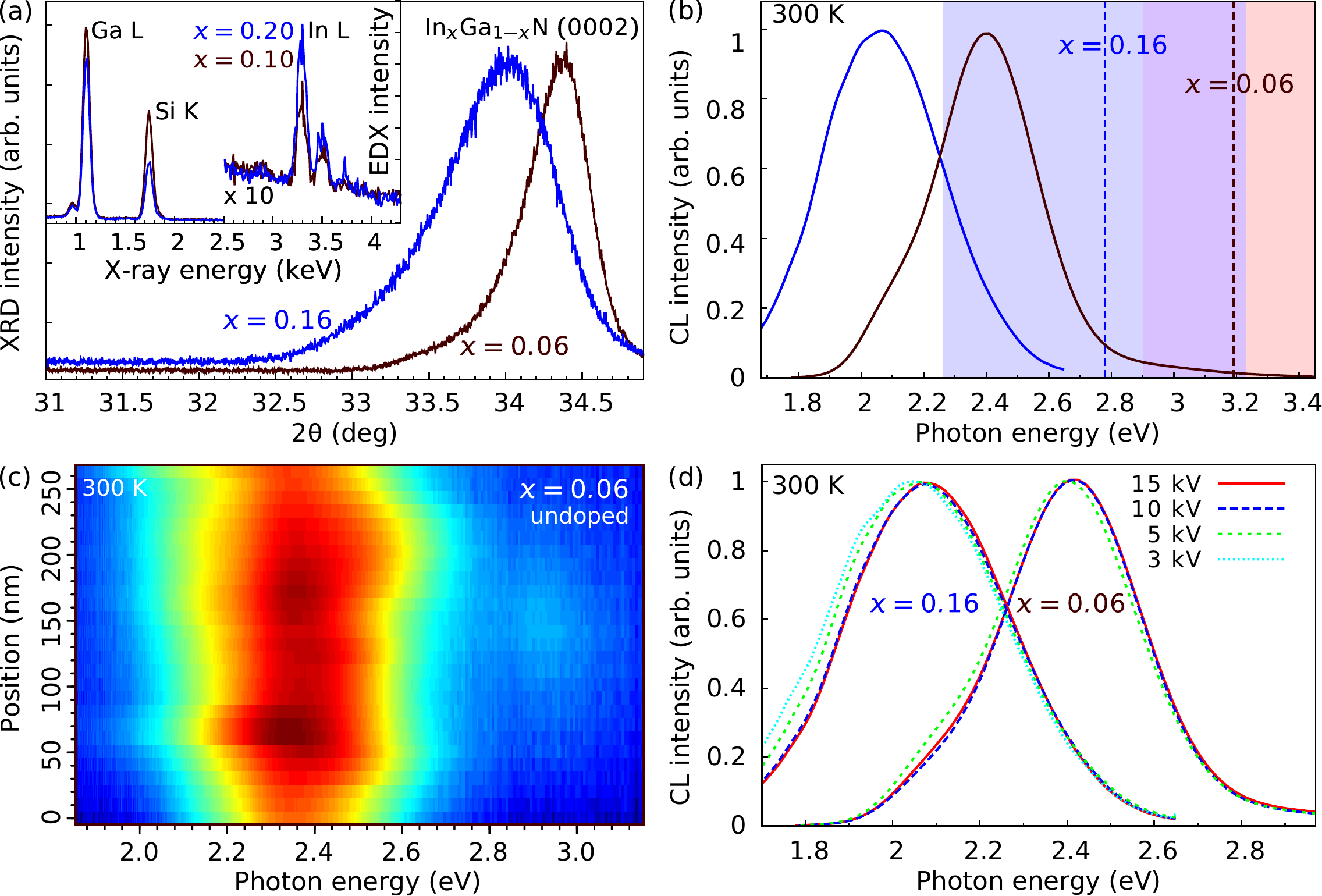}
\caption{\label{fig:xrd+cl}(a) $\upomega$-$2\uptheta$ x-ray diffraction scans of the undoped (In,Ga)N nanowire ensembles on Si(111). Average compositions of $x=0.06$ and $0.16$ are determined from the respective peaks of the (0002) reflection profiles. The inset shows energy-dispersive x-ray spectra of the Ga L, the Si K, and the In L lines for the two samples giving average compositions of $x=0.1$ and $0.2$. (b) Normalized cathodoluminescence spectra of the undoped nanowire ensembles. The dashed lines mark the peak energies expected for bulk material with the given compositions, whereas the shaded areas give the energy range corresponding to the linewidth of the x-ray diffraction profiles. (c) Cathodoluminescence spectral line scan (intensity color coded on a logarithmic scale) along the axis of a single (In,Ga)N nanowire with $x = 0.06$, representative for a number of such scans on both undoped samples. (d) Cathodoluminescence spectra acquired on the undoped nanowire ensembles in top view for different acceleration voltages $V_\text{acc}$.}
\end{figure*}

To give an overview of the samples under investigation, Fig.~\ref{fig:sem} shows top-view SEM images taken on two unintentionally doped (hereafter referred to as \emph{undoped}) nanowire ensembles grown by molecular beam epitaxy at substrate temperatures $T_\text{sub}$ of 590 and 640\,\celsius and from two Si-doped samples with $T_\text{sub} = 640$\,\celsius and different doping levels. For each nanowire ensemble, the distribution of equivalent disk diameters\cite{Brandt2014} is shown in the corresponding inset. As given in Table~\ref{Table1}, the nanowires are about 270~nm long, and their average diameter increases when $T_\text{sub}$ is reduced. This evolution agrees with the results of Refs.~\citenum{Goodman2011} and \citenum{Kamimura2014} showing that for low substrate temperatures nanowires tend to grow laterally. Lateral growth also results in the coalescence of adjacent nanowires, which is reflected in the pronounced tail of the diameter distribution toward large diameters. The degree of coalescence $\sigma_C$ has been quantified using the definition in Ref~\citenum{Brandt2014}, and the corresponding values are reported in Table~\ref{Table1}. With increasing Si doping level, we observe that both the mean diameter of the nanowires and the coalescence are reduced.

Figure \ref{fig:xrd+cl}(a) displays $\upomega$-$2 \uptheta$ x-ray diffraction scans for the two undoped samples. Assuming that the In$_x$Ga$_{1-x}$N nanowires are entirely relaxed,\cite{Jenichen2011} the peak position of the (0002) reflection indicates that the average In composition $x$ is 0.06 and 0.16 for the nanowire ensembles grown at $T_\text{sub} = 640$ and 590\,\celsius, respectively. For both samples, the (0002) reflection is broad and asymmetric. Based on the linewidth of the (0002) reflection and neglecting the contribution of the shortest nanowires to the broadening, we obtain ranges in  In content between $x = 0$ and 0.13 for $T_\text{sub} = 640$\,\celsius and between $x = 0.05$ and 0.3 for $T_\text{sub} = 590$\,\celsius. As summarized in Table~\ref{Table1}, the average In contents deduced by energy-dispersive x-ray spectroscopy [see spectra in the inset of Fig.~\ref{fig:xrd+cl}(a)] are slightly larger than those obtained by x-ray diffractometry. This finding may arise from the low accuracy of energy dispersive x-ray spectroscopy for a quantitative composition analysis in the case of nanowire samples, as discussed in the methods section.

Cathodoluminescence spectra taken at 300~K on the two undoped (In,Ga)N nanowire ensembles are shown in Fig.~\ref{fig:xrd+cl}(b). The emission from the samples grown at $T_\text{sub} = 640$ and 590\,\celsius is centered at 2.40 and 2.05~eV, respectively. Assuming that the bandgap of In$_x$Ga$_{1-x}$N  is given by $E_{\text{In}_x\text{Ga}_{1-x}\text{N}} = 3.45 (1 - x) + 0.68 x - 1.72 x (1 - x)$,\cite{Schley_prb_2007} the In contents corresponding to the peak energies in Fig.~\ref{fig:xrd+cl}(b) are $x = 0.26$ and 0.36, respectively. In other words, for both samples, the emission energy is about 800~meV lower than expected from the average composition.

\begin{figure*}[t]
\centering
\includegraphics*[width=14cm]{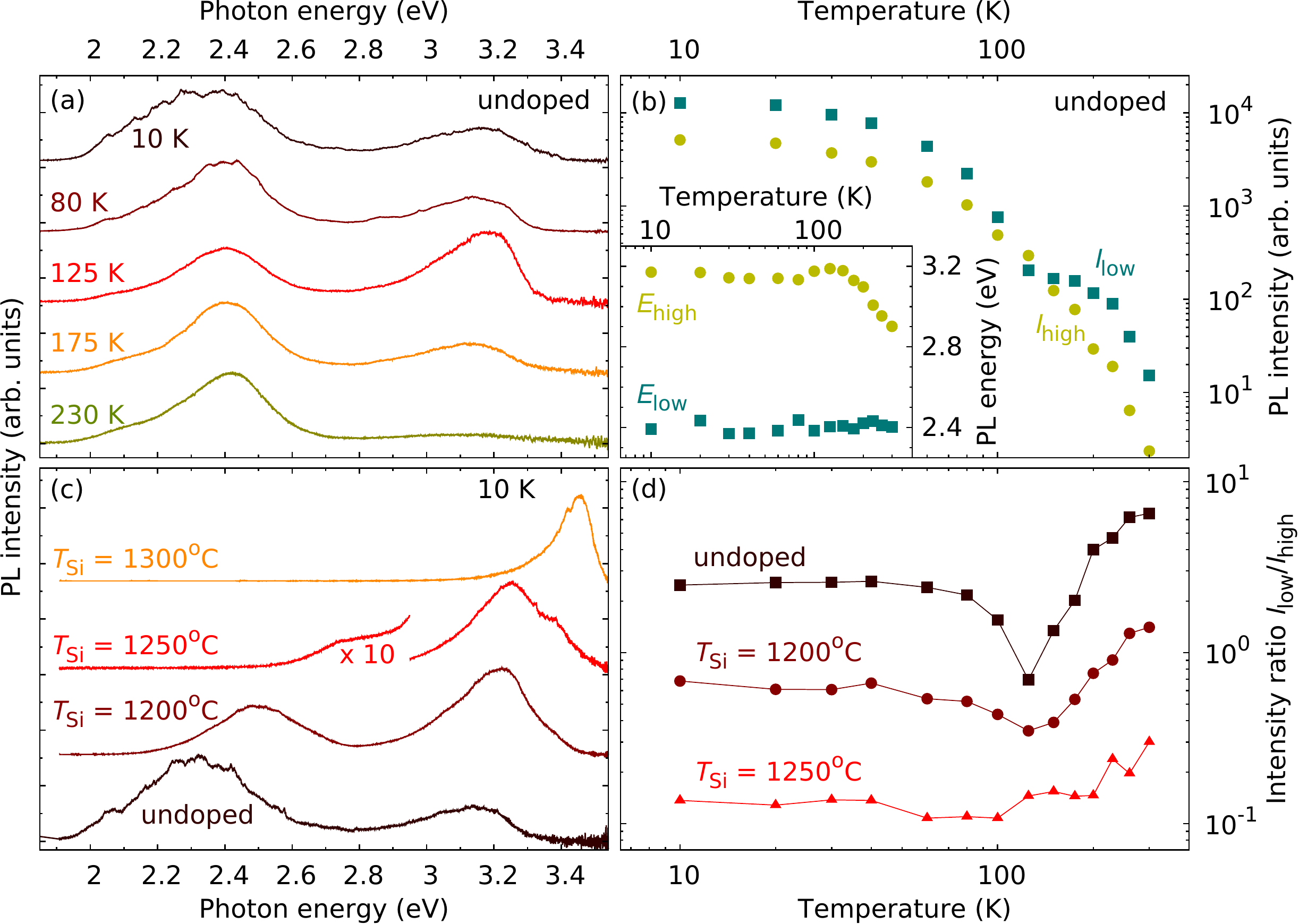}
\caption{\label{fig:pl}(a) Photoluminescence spectra measured at temperatures between 10 and 230~K for the undoped sample with $x = 0.06$. As shown in (b), the spectrally integrated intensities $I_\text{low}$ and $I_\text{high}$ of the two emission bands centered around 2.40 and 3.12~eV at 10~K quench with temperature in distinctly different ways. The inset in (b) shows the evolution of the PL peak energies $E_\text{low}$ and $E_\text{high}$ of the two emission bands with temperature, showing this dependence to be different as well. (c) Photoluminescence spectra measured at 10~K for (In,Ga)N nanowire ensembles for different Si doping levels. (d) Evolution of the intensity ratio $I_\text{low}/I_\text{high}$ as a function of temperature for the different doping levels.}
\end{figure*}

Previous works have demonstrated the presence of pronounced In composition gradients along the axis of (In,Ga)N nanowires.\cite{Segura-Ruiz2011,Segura-Ruiz2014} Such a gradient could potentially contribute to the discrepancy between emission energy and average In content. Figure~\ref{fig:xrd+cl}(c) shows a typical cathodoluminescence scan obtained at 300~K on a nanowire from the undoped ensemble for $T_\text{sub} = 640$\,\celsius. We do not observe any energy shift along the nanowire length. The spatial resolution of the cathodoluminescence experiments is governed by the spatial extent of the generation volume and by the minority carrier diffusion length. The carrier diffusion length in GaN and in In$_x$Ga$_{1-x}$N with low $x$ values is below 100~nm,\cite{Sonderegger2006,Corfdir2011,Nogues2014} and the spatial extent of the generation volume for an acceleration voltage $V_\text{acc} = 5$~kV is much smaller than the nanowire length.\cite{Parish2006} Hence, the absence of any spectral shift in Fig.~\ref{fig:xrd+cl}(c) rules out any significant compositional gradient along the axis of the nanowires under investigation. This finding is confirmed on a larger number of nanowires in Fig.~\ref{fig:xrd+cl}(d), which shows the evolution of the cathodoluminescence spectra in top-view geometry with varying $V_\text{acc}$ for the two undoped nanowire ensembles. According to Monte Carlo simulations of the generation volume,\cite{Drouin_scanning_2007} 75\% of the beam energy is deposited down to depths of 30, 70, 210, and 350~nm for $V_\text{acc}=$3, 5, 10, and 15~kV, respectively. At the same time, the number of generated carriers increases linearly with $V_\text{acc}$.\cite{Lappe1961} The generated carrier density thus decreases by a factor of about 5 with $V_\text{acc}$ increasing from 3 to 15~kV. Since the cathodoluminescence spectra show only minor shifts as a function of $V_\text{acc}$, these experiments confirm the absence of a significant compositional gradient along the nanowire axis.

It is well known that planar (In,Ga)N layers exhibit a large Stokes shift that results from compositional inhomogeneities.\cite{Chichibu1997,Martin1999} Similarly, in (In,Ga)N-based nanowire heterostructures, the combination of compositional fluctuations\cite{Goodman2011,Segura-Ruiz2011,Murotani2013,You2013,Segura-Ruiz2014,Lahnemann_jpd_2014} and inhomogeneous strain\cite{Tourbot2011} can induce a strong redshift of the nanowire emission. In addition, (In,Ga)N nanowires may exhibit large densities of stacking faults,\cite{Tabata2013} which induce charge carrier localization along the nanowire axis,\cite{Graham2013,Lahnemann2014} and the random distribution of dopants in these nanoscale structures can also localize charge carriers.\cite{Marquardt2015}

To get a deeper insight into the localization and recombination dynamics of charge carriers in our nanowires, we have carried out temperature-dependent photoluminescence experiments. Figure~\ref{fig:pl}(a) displays the photoluminescence spectra taken between 10 and 230~K on the undoped nanowire ensemble for $T_\text{sub} = 640$\,\celsius. Similar experiments carried out on the undoped nanowires grown at $T_\text{sub} = 590$\,°C are shown in the Supplementary Information (Fig.~S2). At 10~K, the spectrum is dominated by a band centered at 2.40~eV. In addition, a weaker emission band is observed at 3.12~eV. Two distinct emission bands at low temperatures have also been observed in Ref.~\citenum{Albert2013}. These two bands exhibit a significant spectral overlap, as can be noted when plotting the spectrum on a semi-logarithmic scale (see Supplementary Information). The bands at 2.40 and 3.12~eV are both inhomogeneously broadened and exhibit a full width at half maximum of 0.35 and 0.22~eV, respectively. While the energy of the low-energy band corresponds fairly well to the peak emission energy measured at 300~K in cathodoluminescence spectra, the emission energy of the high-energy band coincides with the (In,Ga)N bandgap calculated using the In content obtained by x-ray diffractometry [Fig.~\ref{fig:xrd+cl}(b)]. In addition, the inset in Fig.~\ref{fig:pl}(b) shows that the energy of the 2.40~eV band remains almost constant between 10 and 300~K, while the high-energy band is redshifted at high temperatures. Neither of these bands exhibit a temperature dependence compatible with the temperature dependence of the band gap.

It is thus tempting to ascribe the presence of these two bands at 3.12 and 2.40~eV to the recombination of carriers bound to shallow and deep potential fluctuations, respectively. To test this hypothesis, we study the evolution of the emission intensity of these two lines as a function of temperature: an increase in temperature should lead to a redistribution of charge carriers from deeply to weakly localized states,\cite{Chichibu1997,Li2001} and is thus expected to give rise to a quenching of the low-energy emission to the benefit of the high-energy band.

Figure~\ref{fig:pl}(b) displays the temperature dependence of the emission intensities of the high- and low-energy bands ($I_\text{high}$ and $I_\text{low}$, respectively), and the ratio $I_\text{low}/I_\text{high}$ is shown in Fig.~\ref{fig:pl}(d). Clearly, the evolution of the photoluminescence spectra as a function of temperature is inconsistent with the assignment of the two emission bands to shallow and deep potential fluctuations. The total nanowire emission intensity decreases with temperature, indicating the activation of a nonradiative recombination channel related either to point defects\cite{Hauswald2014} and/or to the nanowire surface.\cite{Guo2010,Nguyen2012} However, $I_\text{low}$ and $I_\text{high}$ do not exhibit the same dependence with temperature. While $I_\text{high}$ decreases continuously between 10 and 300~K, $I_\text{low}$ decreases between 10 and 100~K, remains nearly constant between 100 and 150~K and decreases again for higher temperatures. The ratio $I_\text{low}/I_\text{high}$ features a pronounced minimum at a temperature $T_\text{min} = 125$~K, and the temperature dependence of $I_\text{low}/I_\text{high}$ cannot be fit using an Arrhenius behavior (a qualitatively similar behavior was observed for the sample with $x=0.16$, as shown in the Supplementary Information). This finding is in contrast to what is expected for the delocalization of carriers from strongly to weakly localized states. Consequently, localization at compositional fluctuations or dopants alone cannot account for the temperature dependence of the emission spectra of our (In,Ga)N nanowires.

Several additional observations allow us to conclusively rule out In-rich clusters as the origin of the low-energy band. First, the density of these clusters needs to be small enough not to be detected by XRD [Fig.~\ref{fig:xrd+cl}(a)]. Therefore, the transitions related to In-rich clusters should be easily saturated when using a high and localized excitation as the one for the cathodoluminescence experiments, in contradiction with what is seen experimentally in Fig.~\ref{fig:xrd+cl}(d). Second, as shown in Ref.~\citenum{Kamimura2013}, our (In,Ga)N nanowires exhibit a broadband absorption. To get such a broadband absorption, one requires a high density of In-rich clusters that should be detected in XRD. Since this is not the case [Fig.~\ref{fig:xrd+cl}(a)], we rule out the presence of such clusters in our nanowires.

\begin{figure}[t!]
\includegraphics*[width=8.5cm]{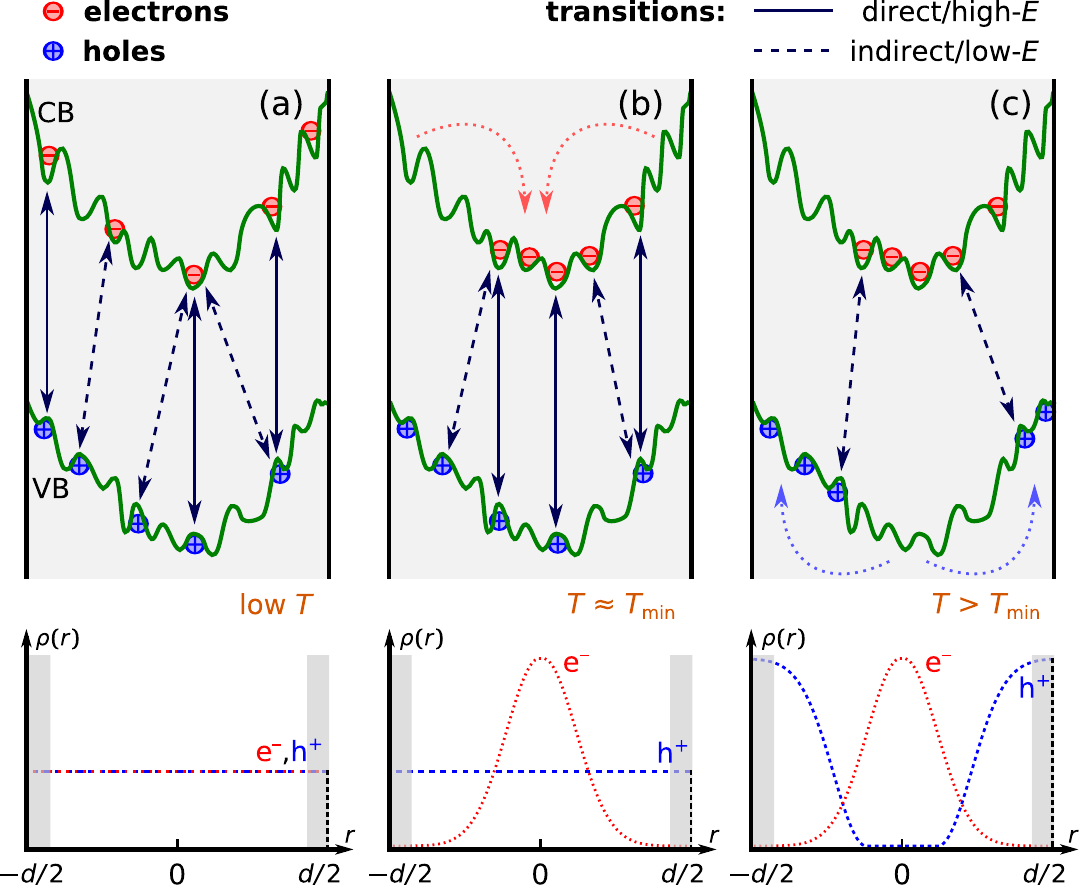}
\caption{\label{fig:bandwiggle}The upper part shows sketches of the lateral band profile across an (In,Ga)N nanowire, where the long-range depletion field is superimposed by potential fluctuations in the ternary alloy. Electrons in the conduction band (CB) and holes in the valence band (VB) can be localized in separate potential minima in radial direction. As a consequence of the depletion fields, a radial Stark shift reduces the transition energy of radially indirect transitions (dashed double arrows) compared to the one of direct transitions (solid double arrows). The three panels represent the situation (a) at low temperature, (b) at the intermediate temperature $T_\text{min}$ where electrons move toward the center of the nanowire, favoring direct transitions, and (c) at a temperature sufficiently high for holes to move towards the sidewalls of the nanowire and for the radially indirect transitions to dominate the photoluminescence spectrum. The lower graphs sketch the corresponding mean probability densities $\rho(r)$ for the distribution of the electrons and holes across the nanowire.}
\end{figure}

To understand the evolution of the emission spectra with temperature as depicted in Figs.~\ref{fig:pl}(b) and \ref{fig:pl}(d), we also have to consider the pinning of the Fermi level at the sidewalls of the nanowires, which may strongly modify their optical properties.\cite{vanWeert2006,Marquardt2013,Corfdir2014} In GaN, the Fermi level is pinned approximately 0.6~eV below the minimum of the conduction band, and unintentionally doped nanowires with a diameter below 100~nm are fully depleted.\cite{Calarco2005,Segev2006,Calarco2011,Schuster2015} We expect that this experimental result applies as well to our unintentionally doped (In,Ga)N nanowires, since they have a low In content, an average diameter of less than 115~nm (Table~\ref{Table1}), and a background doping level that is likely to be on the order of or larger than $10^{17}$~cm$^{-3}$.\citep{Calarco2005,Kamimura2014} The depletion induces radial built-in electric fields that pull the electron and hole wavefunctions toward the core and the sidewalls of the nanowire, respectively. For GaN nanowires, this separation of electrons and holes results in a total quenching of luminescence if the fields exceed a magnitude sufficient for the dissociation of excitons.\citep{Corfdir2014}

For the ternary alloy (In,Ga)N, however, the localization of electrons and holes is not only determined by the radial electric fields, but also by compositional fluctuations. These fluctuations may induce the localization of excitons or, as reported by several groups, the individual localization of electrons and holes.\cite{Morel2003,Brosseau2010,Lahnemann2011,Cardin2013}
The radial spatial separation of electrons and holes may then cause transitions with arbitrary redshift (up to the energy of the pinned Fermi level) compared to the actual bandgap of the material. Figure~\ref{fig:bandwiggle} illustrates this idea, schematically representing the potential landscape for electrons and holes in the nanowires accounting for the radial built-in electric fields and the potential fluctuations that arise from inhomogeneities in the alloy composition and dopant distribution. In such a complex potential landscape, electrons and holes may localize at random positions and, in particular, independently. The spatial distribution of electrons and holes and the corresponding recombination energy may then depend intimately on the magnitude of the radial electric fields, on the depth of the local potential fluctuations, and on temperature.

The observation at low temperatures of a two-band behavior instead of a single and broad emission band may at first appear surprising. To explain this finding, let us assume that (i) potential fluctuations follow a Gaussian distribution with a mean value $E_0$ sufficiently deep to localize holes, and (ii) that the energy depth $E_T$ required to localize electrons is much larger than $E_0$. It follows directly that the density of energy minima capable of localizing an electron and a hole at the same site is small. Since built-in electric fields are stronger close to the surface, the tunneling rate is larger for localized states located close to the sidewalls of the nanowires. Since the density of energy minima deep enough to localize electrons is small, an electron tunneling/hopping out of a potential minimum close to the surface is unlikely to localize in the immediate vicinity, but most likely proceeds towards the center of the nanowire. Consequently, at low temperature, electrons and holes are either located at the same site, giving rise to the high energy band, or exhibit a large radial separation, giving rise to the low energy band, and intermediate emission energies are unlikely, in agreement with what is observed experimentally in Fig.~\ref{fig:pl}. Note that the characteristic energies $E_0$ and $E_T$ and thus $I_\text{low}/I_\text{high}$ depend on the magnitude of the radial fields at the surface and on the depth of the potential fluctations. Consequently, the exact shape of the photoluminescence spectrum, as well as the ratio $I_\text{low}/I_\text{high}$, vary in a complex fashion with $T_\text{sub}$, $T_\text{Si}$, and $\sigma_C$.

Spatially direct transitions, which correspond to the high energy emission band, seem to be much favored over indirect ones because of their higher overlap and thus shorter radiative lifetime. However, the separation of electrons and holes not only slows down their radiative recombination, but also nonradiative processes: any Shockley-Read-Hall recombination event requires the interaction of both electron and hole with the recombination center. The combined overlap of separately localized electrons and holes with nonradiative recombination centers is just as low as their mutual overlap. For statistically localized electron-hole pairs, the nonradiative rate is thus decreased concurrently with the radiative one.

To test these ideas, we analyze the ratio $I_\text{low}/I_\text{high}$, which is a function of the recombination rates of direct and indirect transitions. Since the effective mass of electrons is significantly lower than that of holes, an increase in temperature leads first to an increased transfer rate of electrons toward the center of the nanowire, as indicated in Fig.~\ref{fig:bandwiggle}(b), while the transfer hole remain mostly frozen. With  more electrons becoming available for direct transitions in the core of the nanowire, this recombination channel is promoted, leading to the decrease in $I_\text{low}/I_\text{high}$ as observed in Fig.~\ref{fig:pl}(d) when the temperature approaches $T_\text{min}$. However, when the temperature increases further, also holes can escape the potential fluctuations and drift toward the surface of the nanowire, as displayed in Fig.~\ref{fig:bandwiggle}(c). The average spatial separation between the electron and hole wavefunctions is then increased again, decreasing the proportion of the spatially direct high-energy transitions and increasing that of the low-energy transition. This development leads to an increase in $I_\text{low}/I_\text{high}$. At 300~K, we find that  $I_\text{low} \gg I_\text{high}$, implying that the surface band bending is larger than $k_B T$ and that the relaxation of electrons and holes in our (In,Ga)N nanowires is faster than the radiative lifetime for the spatially direct recombination.

\begin{figure}[t!]
\centering
\includegraphics*[width=\columnwidth]{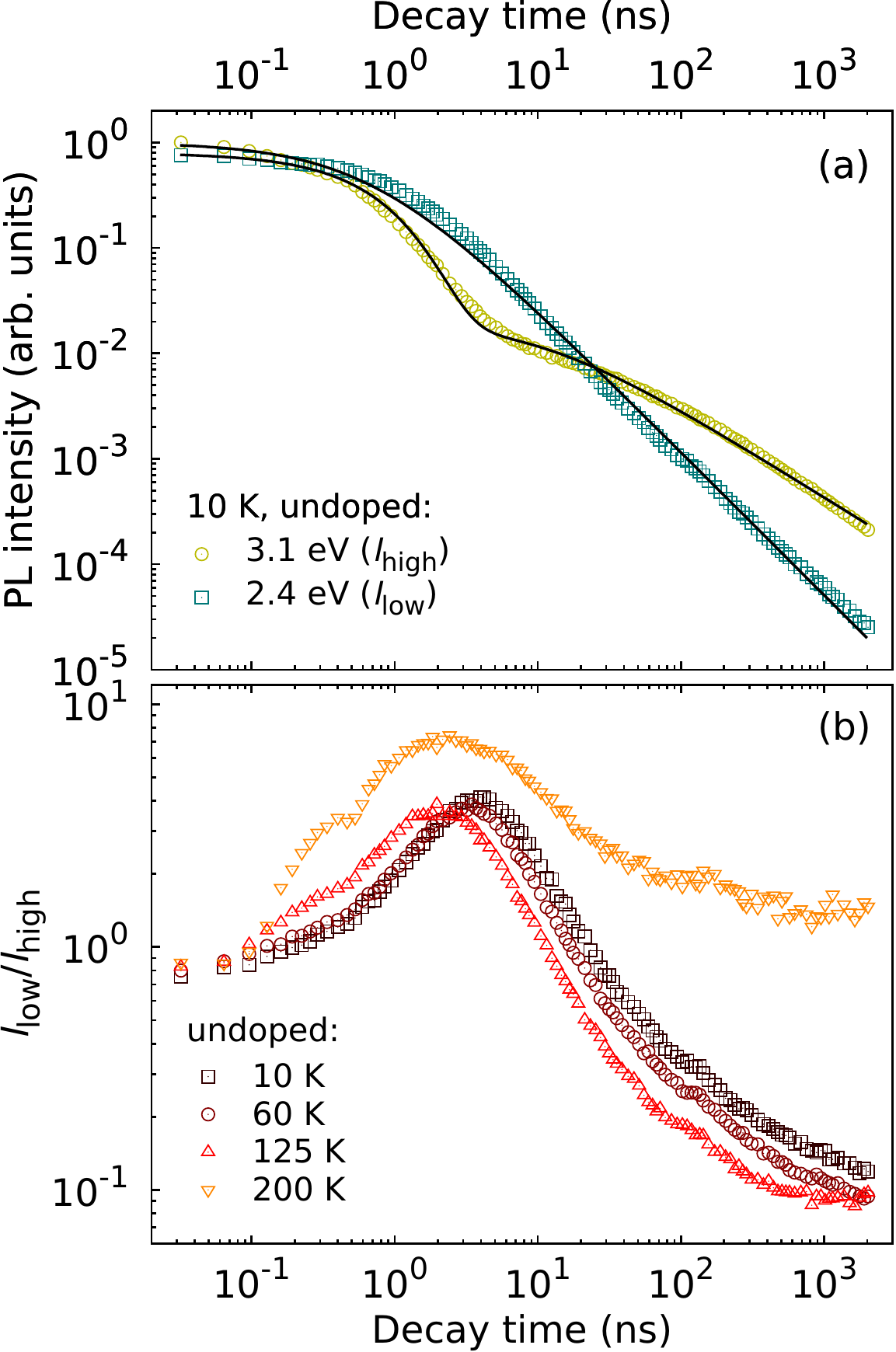}
\caption{\label{fig:trpl}(a) Photoluminescence transients at 2.4 and 3.1~eV recorded at 10~K for the undoped sample with $T_\text{sub} = 640$\,\celsius plotted on a double-logarithmic scale. Noticeable is the slowdown of the decay of $I_\text{high}$ for a delay of 4~ns. The solid lines show fits to the data as discussed in the text. (b) Temporal evolution of the ratio $I_\text{low}/I_\text{high}$ measured at different temperatures.}
\end{figure}

As an independent test for the attribution of the high- and low-energy transitions to spatially direct and indirect transitions, respectively, we investigate samples with various doping levels. As shown in Refs.~\citenum{Calarco2005}, \citenum{Calarco2011}, and \citenum{Schuster2015}, the depletion depth within the nanowires can be controlled by doping. For unintentionally doped (In,Ga)N with a residual $n$-type doping in the mid $10^{17}$~cm$^{-3}$, the entire nanowire interior experiences a strong built-in electric field. For higher doping densities, the fraction of the nanowire volume experiencing strong radial fields is reduced, which should lead to a decrease of $I_\text{low}/I_\text{high}$. Figure~\ref{fig:pl}(c) shows photoluminescence spectra at 10~K for (In,Ga)N nanowire ensembles with different doping levels induced by intentional Si doping. The two emission bands are observed up to a Si cell temperature of 1250\,\celsius. With increasing Si doping concentration, the high-energy band blueshifts by 0.3~eV, in agreement with the reduction in In content noticed in Table~\ref{Table1}. The significant blueshift for $T_\text{Si}$ larger than 1250\,°C may also result from a change in strain state originating from the decrease in $\sigma_C$. The energy shift for the low-energy band is larger ($\approx 0.6$~eV), indicating that for higher Si concentrations the increasing electron density in the nanowire leads to a partial screening of the surface states and thus to a reduction of the energy at which the Fermi level is pinned.\cite{Schuster2015} Most importantly, the ratio $I_\text{low}/I_\text{high}$ decreases from 2.5 to 0.14 with increasing Si incorporation, and the low-energy band vanishes for the highest doping concentration ($T_\text{Si}=$1300\,\celsius). This finding is in complete agreement with the one explained above based on the interpretation of the low-energy transition as a spatially indirect one.

A fingerprint for the individual localization of electrons and holes as depicted in Fig.~\ref{fig:bandwiggle} is the power law decay of the emission intensity after pulsed excitation.\cite{Morel2003,Brosseau2010,Lahnemann2011,Cardin2013,Sabelfeld2015} Figure~\ref{fig:trpl}(a) shows the time evolution of the low- and high-energy emission bands at 10~K after pulsed excitation. The transients have been integrated spectrally integrated. Note that there is no significant spectral dependence of the transients within the respective emission bands (see Supplementary Information). The transient attributed to the spatially indirect transitions closely resembles the decay reported for (In,Ga)N/GaN quantum wells in Ref.~\citenum{Brosseau2010} and can be fit fairly well by a simple phenomenological power law\cite{Cardin2013} as shown in Fig.~\ref{fig:trpl}(a). This result confirms that alloy disorder in our (In,Ga)N nanowires leads to an independent localization of electrons and holes. The time dependence of the high-energy transition is more complex and is characterized by an initial rapid decay, which slows down considerably after 4~ns. For the longest times, the decay is even slower than that of the low-energy transition. A close inspection of this transient reveals that the initial decay is in fact close to a single exponential with a lifetime of about 700~ps, while it is clearly non-exponential at times longer than 4~ns. We found that the transient can be described very well [cf.\ Fig.~\ref{fig:trpl}(a)] by a simple sum of an exponential and the same phenomenological power law as used for the low-energy transition, albeit with different parameters. This result shows that the high-energy emission is actually a superposition of spatially direct transitions having high oscillator strength (and thus short radiative lifetime) and spatially indirect transitions occurring at the same energy, i.\,e., indirect transitions that do \emph{not} exhibit any Stark shift. Such transitions are possible for electrons and holes separated not radially, but vertically along the nanowire axis.   

This superposition of short- and long-lived excitations is most clearly reflected by the non-monotonic time dependence of $I_\text{low}/I_\text{high}$ displayed in Fig.~\ref{fig:trpl}(b). At 10~K, the ratio $I_\text{low}/I_\text{high}$ directly after the pulse is 0.75. It increases up to 4 after about 4~ns and then decreases to a value of 0.1 for longer times. The low-energy transition originates from electrons and holes separated in \emph{radial} direction, resulting in a pronounced Stark shift. The high-energy transition, in contrast, is dominated by spatially direct and most likely excitonic transitions. The rapid decay of these transitions as compared to those constituting the low-energy band leads to the initial increase in $I_\text{low}/I_\text{high}$. At times longer than 4~ns, however, the intensity of these excitonic transitions has decayed by two orders of magnitude, and transitions of electrons and holes separated along the \emph{axial} direction start to dominate at the same energy. The corresponding slowdown of the decay results in the decrease of the ratio $I_\text{low}/I_\text{high}$ for longer delays. With an increase of the temperature from 10 to 200~K, the maximum of $I_\text{low}/I_\text{high}$ shifts to shorter times [Fig.~\ref{fig:trpl}(b)], since the decay of $I_\text{high}$ accelerates, which we attribute to an increasing participation of nonradiative processes in direct transitions. As discussed before, nonradiative recombination is much less pronounced for the indirect transitions. The temperature dependence of $I_\text{low}/I_\text{high}$ for long time delays and for continuous-wave photoluminescence spectroscopy [Fig.~\ref{fig:pl}(d)] show a similar behavior: $I_\text{low}/I_\text{high}$ decreases between 10~K and $T_\text{min} = 125$~K and increases for higher temperatures. 
\begin{figure}[t]
\centering
\includegraphics*[width=\columnwidth]{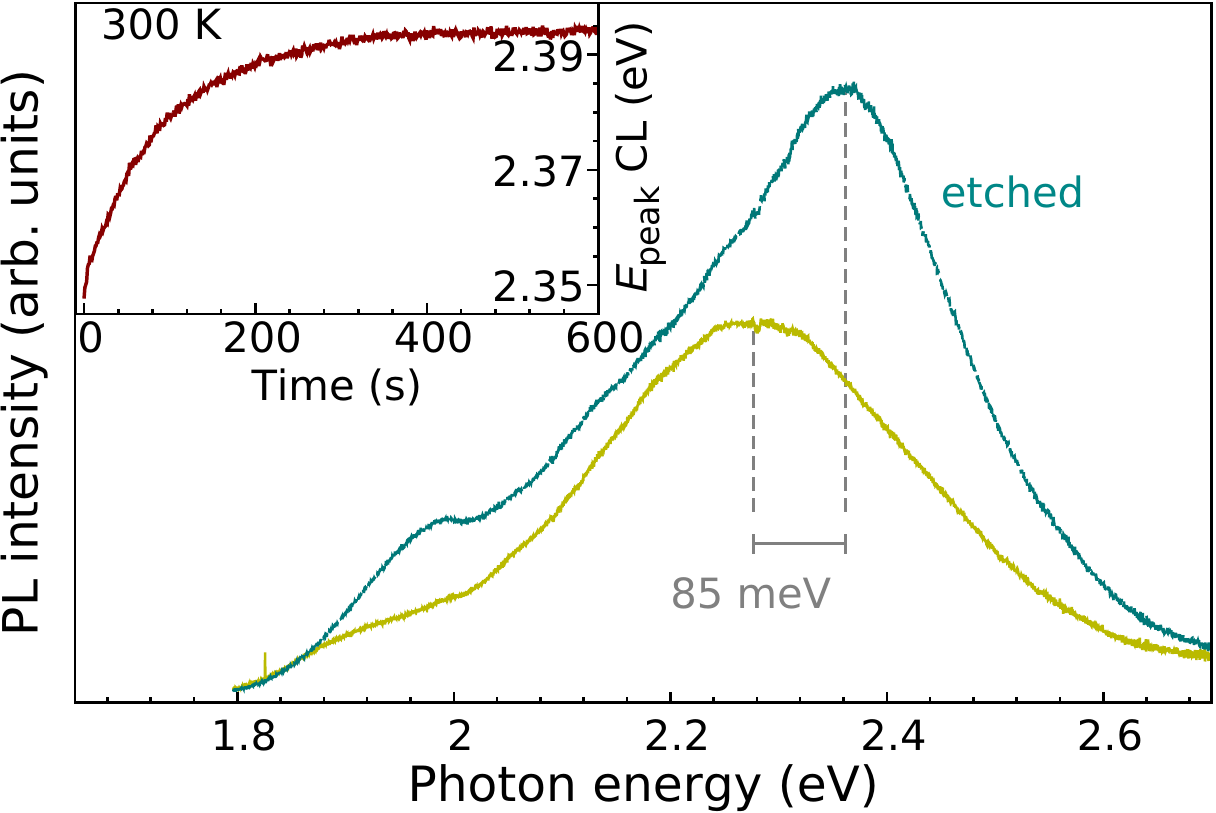}
\caption{\label{fig:cltime}Room temperature photoluminescence spectra recorded on an ensemble of undoped In$_x$Ga$_{1-x}$N nanowires with $x=0.06$ prior to (yellow) and after (cyan) a dilute HCl etch resulting in a blueshift of 85~meV. The inset shows the evolution of the peak energy of the emission band under prolonged electron irradiation as obtained by spectrally-resolved top-view cathodoluminescence measurements on the nanowire ensemble over a period of 10~minutes. During the first 4~min, the peak is blueshifted by about 50~meV.}
\end{figure}

Any modification of the surface states leads to changes in the Fermi level pinning and the surface band bending. Regarding the magnitude of the radial Stark effect discussed above, such a modification should manifest itself by a change in the emission spectrum of the nanowires. Figure~\ref{fig:cltime} shows the photoluminescence spectrum at 300~K from the unintentionally doped nanowire ensemble with $x=0.06$ before and after a treatment with 30\% HCl. As shown in Ref.~\citenum{Gurwitz2011}, hydrochloric acid (HCl) removes the native Ga oxide from the nanowire surface, leading to a reduction of both the surface state density and the surface band bending. The nanowire emission measured right after a 30~s HCl etching exhibits a blueshift of 85~meV. This finding is consistent with a reduction of the radial electric fields across the nanowire affecting the indirect transitions, which dominate the luminescence spectrum  at room temperature. This reduction of the fields should also result in an increase in electron-hole overlap and thus in an improved radiative efficiency, and we indeed observe a significant increase in photoluminescence intensity after the HCl etch.

An alternative way to modify the magnitude of the radial electric field is electron irradiation, which leads to a complex interplay between charge trapping and surface state modification by carbon contamination.\cite{Campo2004,Robins2007} The inset of Fig.~\ref{fig:cltime} shows the evolution of the cathodoluminescence peak energy of the undoped nanowire ensemble at 300~K as a function of the electron beam exposure time. During an exposure of 240~s, the emission band blueshifts by about 50~meV, indicating again the screening of the radial electric fields. Consistent with the reduced influence of surface electric fields with increasing Si doping level observed in the photoluminescence experiments, this pronounced peak shift under electron beam exposure is not observed in the doped samples. Finally, also the results of oxygen photodesorption experiments carried out on (In,Ga)N nanowires are in line with our interpretation. As shown in Ref.~\citenum{Lefebvre2012}, the modification in surface band bending due to the adsorption of oxygen at the sidewalls of (In,Ga)N nanowires only affects the low-energy part of the photoluminescence spectrum, i.\,e., the energy range corresponding to the radially indirect transitions.

In conclusion, we have shown that the combination of surface electric fields and carrier localization strongly affects the emission properties of spontaneously formed (In,Ga)N nanowires with a low In content. In this context, we have highlighted the dominant role of radially indirect transitions between electrons and holes localized close to the core and the surface of the nanowire, respectively, for emission at room temperature. On the one hand, the band bending arising from the pinning of the Fermi level at the sidewalls of the nanowires results in a radial Stark effect that redshifts the emission from radially indirect transitions. On the other hand, compositional inhomogeneities and the random donor distribution localize electrons and holes, hindering their complete spatial separation. At cryogenic temperatures, this competition results in the observation of two broad emission bands essentially related to spatially direct and indirect recombination. We have described in detail the dynamics of these two bands as well as their evolution with temperature. The magnitude of the Stark shift and the intensity of the radially indirect transition depend on the doping concentration and can be modified using surface treatments such as HCl etching or electron irradiation. A similar two-band behavior is expected for any ternary semiconductor alloy grown in the form of nanowires, with a high doping level and a pronounced alloy disorder. In contrast, for binary compounds such as GaN, the photoluminescence at low temperature is dominated by the recombination of excitons bound to neutral donors and acceptors.\cite{Corfdir2014} In this case, the radial Stark effect only leads to a quenching of the photoluminescence intensity at low temperature and no energy shift is observed, as reported in Ref.~\citenum{Pfuller2010}.

The quantum-confined Stark effect is usually seen as a detrimental phenomenon, since it leads to a decrease of radiative efficiency at room temperature. For some applications, however, the quantum-confined Stark effect is beneficial.\citep{Damilano1999} In the present context, it is important to note that the quantum.confined Stark effect shifts the absorption edge toward longer wavelengths.\cite{Miller1985,Chow1999,Kalliakos2003} Recent measurements of the incident-photon-to-current conversion efficiency have shown that (In,Ga)N nanowires with a low In content exhibit absorption throughout the visible spectral range.\cite{Kamimura2013} This broadband absorption, which results from the radial Stark effect discussed in the present paper, makes (In,Ga)N nanowires attractive for solar energy applications while limiting the demand for the relatively scarce element In. In the specific case of solar water splitting, the theoretical solar-to-hydrogen conversion efficiency is maximum and equal to 47\% when the bandgap of the semiconductor working electrode is equal to 1.23 eV and decreases rapidly for larger bandgaps \cite{Chen2013}. The redshift of the bandedge resulting from the radial Stark effect thus leads to an increase in the solar-to-hydrogen conversion efficiency for (In,Ga)N nanowires with low In-content. In particular, the redshift of the bandedge from 3.2 to 2.4 eV observed in Fig.~\ref{fig:xrd+cl} for the nanowires with $x=0.06$ should lead to an increase in the theoretical solar-to-hydrogen conversion efficiency from 1 to 10\%. 

\section{Experimental Methods}\label{ExperimentalDetails}

\paragraph{Growth Details and Composition Analysis}
Nominally undoped and $n$-doped (In,Ga)N nanowires were grown by plasma-assisted molecular beam epitaxy on Si(111) substrates using a self-induced approach.\cite{Kamimura2013,Kamimura2014} The In content was varied by changing the temperature of the substrate ($T_\text{sub}$) between 590 and 640\,\celsius.\cite{Wolz2012} For Si doping, the temperature of the Si effusion cell ($T_\text{Si}$) was varied between 1200 and 1300\,\celsius, corresponding to free carrier concentrations between about 1 and $2 \times 10^{19}$~cm$^{-3}$ at 300~K.\cite{Kamimura2014} Table~\ref{Table1} summarizes the growth temperature, the mean diameter, and the length of the investigated samples.  For each sample, the distribution of the nanowire equivalent disk diameters $d_\text{disk} = 2 \sqrt{A/\pi}$ was obtained by measuring the area $A$ of the nanowire top facets in scanning electron microscopy (SEM) images.\cite{Hauswald2014,Brandt2014} From this distribution, the mean diameter $\langle d_\text{disk} \rangle$ and its variance are deduced by fits with a shifted Gamma distribution.\cite{Hauswald2014} The composition of the (In,Ga)N nanowire ensembles was analyzed by x-ray diffractometry and energy-dispersive x-ray spectroscopy as summarized in Table~\ref{Table1}. Symmetric $\upomega$-$2\uptheta$ x-ray diffraction scans across the (In,Ga)N 0002 reflection were
acquired with Cu$_{K_{\alpha_2}}$ radiation using a Panalytical X’Pert system with a Ge(220) hybrid monochromator and a Ge(220) analyzer crystal. Energy-dispersive x-ray spectroscopy was performed using an EDAX Apollo XV silicon drift detector mounted to a Zeiss Ultra55 field emission scanning electron microscope operating at 7~kV. The compositions obtained by x-ray diffractometry should be more accurate than the ones deduced from energy-dispersive x-ray spectra. The quantification routines used for the analysis of the energy-dispersive x-ray spectroscopy scans have been developed for films that are continuous and homogeneous over the excitation volume. In addition, at the acceleration voltage of 7~kV used to probe the In-L lines, the excitation volume even extends into the Si substrate. However, since the XRD peaks for our (In,Ga)N nanowires are quite broad [see Fig.~\ref{fig:xrd+cl}(a)], small differences in In-content may be difficult to measure reliably using this technique. Therefore, EDX is certainly more suited to evaluate the change in In-content between the samples grown with $T_\text{Si} = 1250$ and 1300\,°C. 

\paragraph{Luminescence Spectroscopy}
The emission properties of the nanowires were studied by photoluminescence and cathodoluminescence spectroscopy. Cathodoluminescence measurements were performed at 300~K in the Zeiss field-emission SEM operating at 5~kV. The nanowire emission was collected using a parabolic mirror and dispersed using a spectrometer (focal length of 30~cm, 300 lines per mm grating) followed by a charge-coupled device (CCD). Continuous-wave photoluminescence experiments were realized using a HeCd laser ($\lambda = 325$~nm) that was focused down to a spot of 1~\textmu m diameter using an objective with a numerical aperture of 0.65. The samples were mounted on a coldfinger cryostat that can reach temperatures between 10 and 300~K. The nanowire emission was collected by a 80~cm focal length spectrometer equipped with a 600 lines per mm grating and was detected with a CCD. Time-resolved photoluminescence experiments were performed at 10~K, using a frequency-doubled fs Ti:sapphire laser (excitation wavelength of 353~nm). A pulse picker was used to reduce the repetition rate to 475~kHz. We estimate the energy fluence per pulse to 10~\textmu J\,cm$^{-2}$. The laser was focused down to a 9~\textmu m diameter spot at the surface of the sample. The photoluminescence signal was detected using a photomultiplier tube, and the synchronization of the photon counting with the excitation was obtained using a time-correlation acquisition system, yielding a time resolution of about 60~ps.

\section{Supporting information}

Additional continuous-wave and time-resolved photoluminescence experiments.

\section{Author contributions}
J.L. and P.C. have contributed equally to this work.

\section{Acknowledgments}
We thank Uwe Jahn for a critical reading of the manuscript. P.\,C. acknowledges partial funding from the Fonds National Suisse de la Recherche Scientifique through project 161032. J.\,K. acknowledges funding from a JSPS Postdoctoral Fellowships for Research Abroad. Part of this work has been funded by the European Commission (FP7-NMP-2013-SMALL-7) under grant agreement no. 604416 (DEEPEN).

\section{Competing financial interests}

The authors declare no competing financial interests.

\bibliography{InGaN-NWs_final}

\end{document}